\documentstyle[12pt,psfig,draft]{nature-ejb}

\def\simlt{\mathrel{\hbox{\rlap{\hbox{\lower4pt\hbox{$\sim$}}}\hbox{$<$}}}}
\def\simgt{\mathrel{\hbox{\rlap{\hbox{\lower4pt\hbox{$\sim$}}}\hbox{$>$}}}}
\newcommand\lsim{\mathrel{\spose{\lower 3pt\hbox{$\mathchar"218$}}
     \raise 2.0pt\hbox{$\mathchar"13C$}}}
\newcommand\gsim{\mathrel{\spose{\lower 3pt\hbox{$\mathchar"218$}}
     \raise 2.0pt\hbox{$\mathchar"13E$}}}

\def\ale{\mathrel{\hbox{\rlap{\hbox{\lower4pt\hbox{$\sim$}}}\hbox{$<$}}}}
\def\age{\mathrel{\hbox{\rlap{\hbox{\lower4pt\hbox{$\sim$}}}\hbox{$>$}}}}

\def\grb{GRB\,031203}
\def\ibis{{\sl IBIS}}
\def\isgri{{\sl ISGRI}}
\def\integral{{\sl INTEGRAL}}

\begin{document}

\title{\Large\bf An apparently normal $\gamma$-ray burst with an unusually low
luminosity}

\author{
S. Yu. Sazonov\affiliationmark[1]\affiliationmark[2]
A. A. Lutovinov\affiliationmark[1] \&
R. A. Sunyaev\affiliationmark[1]\affiliationmark[2] 
\affiliationtext[1]{Space Research Institute, Russian
Academy of Sciences, Profsoyuznaya 84/32, 117997 Moscow, Russia}
\affiliationtext[2]{Max-Planck-Institut f\"ur Astrophysik,
Karl-Schwarzschild-Str. 1, D-85740 Garching bei M\"unchen, Germany}
}

\date{today}{}
\headertitle{An unusually low luminosity GRB}
\mainauthor{Sazonov, Lutovinov \&\ Sunyaev}

\summary{
Much of progress in gamma-ray bursts (GRBs) has come from the studies
of distant events (redshift $z\sim 1$).  The brightest GRBs are
the most collimated events and seen across the Universe due to their
brilliance. It has long been suspected that nearest (and most common)
events have been missed because they are not so collimated or
under-energetic or both \cite{mwh01}.  Here we report soft
$\gamma$-ray observations of \grb, the nearest event to date
($z=0.106$; ref.~\pcite{pbc+04}). 
This event with a duration of 40\,s and peak energy of $>190\,$keV
appears to be a typical long duration GRB. However, the isotropic
$\gamma$-ray energy $\simlt 10^{50}\,$erg, about three orders of
magnitude smaller than the cosmological population. This event as
well as the other nearby but somewhat controversial event GRB\,980425
are clear outliers for the much  discussed isotropic-energy
peak-energy relation \cite{aft+02,ldg04} and luminosity
spectral-lag relations \cite{nmb00,sdb01}. Radio calorimetry shows
that both these events are under-energetic explosions
\cite{skb+04}. We conclude that there does indeed exist a large
population of under-energetic events. 
}

\maketitle

On 2003 December 3 at 22:01:28 UTC, \ibis, a hard X-ray coded
aperture mask imager on the \integral\ satellite detected\cite{mg03}
a pulse of 40\,s-duration. This event, \grb, was localized by
on-board software and the 2.5-arcmin position rapidly
disseminated\cite{gmb+03}. The event with a simple profile
(Figure~\ref{fig:LightCurve}) appears to be a typical long duration GRB. 
Likewise the spectrum is also typical (Figure~\ref{fig:Spectrum}). A
single power law model with photon index $\alpha=-1.63\pm 0.06$
provides an adequate fit.  We place a lower limit on the spectral peak
energy, $E_{\rm peak}>190\,$keV (90\% confidence; see
Figure~\ref{fig:Spectrum}). 

We found no evidence for significant spectral evolution on short
(seconds) time scales. Next, we cross-correlated the light curves
in two energy ranges (soft, 20--50\,keV and hard, 100--200\,keV)
and detected a marginal lag of $0.24\pm 0.12\,$ in the usual sense
(harder emission preceding the softer emission).

Watson et al.\cite{whl+04} have suggested that \grb\ is an X-ray
flash (XRF). XRFs are defined\cite{hzk+01,bol+03} either by $E_{\rm
peak}<50$\,keV or a larger 2--30\,keV fluence ($S_X$) compared to
that in the traditional GRB band, 30--400\,keV ($S_\gamma$). The 
high-$E_{\rm peak}$ soft $\gamma$-ray spectrum measured by \integral\ provides 
direct evidence that \grb\ is a  GRB rather than an XRF.  
The case for an XRF was made by a very high value of flux at 1\,keV, 
$F_X= (2.6\pm 1.3)\times 10^{-6}\,{\rm erg\,cm}^{-2}\,{\rm keV}^{-1}$, 
inferred from modelling of a dust-scattered echo\cite{vwo+04}.
As can be seen from Figure~\ref{fig:Spectrum} the soft X-ray emission
is well above an extrapolation of the \integral\ spectrum (which
predicts $0.36<S_X/S_\gamma< 0.53$, depending on the precise value of
$E_{\rm peak}$) and thus it must have an origin different from the process
producing the soft $\gamma$-ray pulse. It is possible that the echo was 
caused by the early ($\simlt1,000$\,s) afterglow. We note however 
that the inferred soft X-ray fluence is inversely proportional to the 
assumed dust column and extrapolation of our spectrum to keV energies is 
consistent with the lower soft X-ray fluence (due to a higher dust column) 
advocated by Prochaska et al.\cite{pbc+04} 

The burst fluence in the 20--200\,keV band is $(2.0\pm 0.4)\times
10^{-6}$\,erg\,cm$^{-2}$. Adopting the redshift of 0.1
(ref. \pcite{pbc+04}) and the currently popular cosmological parameters 
[($H_0, \Omega_{\rm m}, \Omega_\Lambda)=(75$\,km\,s$^{-1}$\,Mpc$^{-1}$,
0.3, 0.7)].  we find that the isotropic energy equivalent is $(4\pm
1)\times 10^{49}$\,erg. Defining $\epsilon_{\gamma,{\rm iso}}$
to be the isotropic energy equivalent over the 20--2000\,keV band, we
find $6\times 10^{49}\,{\rm erg}<\epsilon_{\gamma,{\rm iso}}<
1.4\times 10^{50}\,{\rm erg}$; where the range reflects the
observational uncertainty in the spectrum above 200\,keV (see
Figure~\ref{fig:Spectrum}).

\grb, an event with spectrum similar to cosmological GRBs, is the
least energetic (in terms of $\epsilon_{\gamma,{\rm iso}}$) long
duration GRB.  Clearly, $\gamma$-ray luminosities and energy releases
vary widely, spanning at least four orders of magnitude. 
Furthermore it is of considerable interest to note that \grb\ violates two
much discussed relations in GRB astrophysics: (1) the
$\epsilon_{\gamma,{\rm iso}}$-$E_{\rm peak}$ relation
and (2) the luminosity-spectral lag relation.  For \grb, the first relation 
predicts\cite{ldg04} $E_{\rm peak}\sim 10$\,keV, in gross 
disagreement with the analysis presented here.  Long spectral lags are
expected from the second relation\cite{nmb00,sdb01} when in fact we see 
virtually no lag (Figure~\ref{fig:Lag-Luminosity}).

\grb, however,  shares some properties with GRB\,980425 associated
with a nearby ($z=0.0085$)  SN~1998bw\cite{gvv+98,kfw+98}.  This
event was also severely underenergetic, $\epsilon_{\gamma,{\rm iso}}\sim
10^{48}\,$erg and violated the $\epsilon_{\gamma,{\rm iso}}$-$E_{\rm
peak}$ relation. Curiously, GRB\,98425 was also a
single pulse\cite{bkh+98} but without a cusp. 

To summarize, the two nearest long duration events,
\grb\ and GRB\,980425 are
clearly sub-energetic in the $\gamma$-ray band. Their proximity
(and hence implied abundance) makes it of prime interest to understand
their origin and relation to the more distant cosmological events.
Are these events genuinely low energy explosions\cite{kfw+98}
(``sub-energetic''  model) or a typical GRB viewed away from its
axis (``off-axis'' model)?

In the off-axis model\cite{yyn03}, $\epsilon_{\gamma,{\rm iso}}\propto
\delta^n$ with $n\sim$2--3 where
$\delta=\gamma^{-1}[1-\beta\cos(\theta-\theta_j)]^{-1}$ is the
so-called Doppler factor; here, $v=c\beta$ is the velocity 
of the shocked ejecta, $c$ is the velocity of light,
$\gamma=(1-\beta^2)^{-1/2}$, $\theta$ is the angle between the
observer and the principal axis of the explosion and $\theta_j$ is the
opening angle of the explosion (``jet''). If we wish to make \grb\ to
have isotropic energies similar to cosmological GRBs then  $\delta$
should be $\sim 10$ to 30 times smaller than the on-axis value
$\delta_0$. The true peak energy is then
$(\delta_0/\delta)\times(1+z)\times E_{\rm peak}> 2\,$MeV -- making
\grb\ one of the hardest bursts. A second consequence is that  
the afterglow should brighten as the ejecta slows down.  Soderberg et
al. \cite{skb+04} do not see any rebrightening and furthermore the
afterglow is faint indicating that the explosion was underenergetic,
$\simlt 10^{50}\,$erg. Likewise, radio calorimetry of SN~1998bw at
early\cite{kfw+98} or late\cite{sfw04} find $\simlt 10^{50}\,$erg.
Thus we conclude that \grb\ and GRB\,980425 are intrinsically
sub-energetic events.

With peak count rate of $1.3$\,photon\,cm$^{-2}$\,s$^{-1}$
(50--300\,keV), \grb\ could have been  detected by {\sl BATSE} out $z\sim
0.25$. So as not to significantly distort the observed (nearly flat)
burst intensity distribution at low fluxes, less than $\sim 300$
underluminous bursts like \grb\ can be present in the {\sl BATSE}
catalogue\cite{pmp+99}, including up to $\sim 20$  located at
$z<0.1$. On the other hand, a ``typical'' GRB with
$\epsilon_{\gamma,{\rm iso}}\sim 10^{53}$\,erg would have a fluence of
$10^{-3}$\,erg\,cm$^{-2}$ if it occured as close as \grb. 
Only a few such bright bursts have been observed in the $\sim
30$\,years of GRB observations\cite{mg81,kst+86,feh+93} suggesting
a large population of events like \grb\ (see also
ref. \pcite{pbc+04}). We eagerly await the launch of the {\sl Swift}
mission which with its increased sensitivity should detect and 
localize many ($\times 10$ the current rate) under-energetic events like \grb.

\bibliographystyle{nature-pap}
\bibliography{journals,refs-Integral1}

\noindent Correspondence and requests for materials should be addressed to
S. Yu. Sazonov (e-mail: sazonov@mpa-garching.mpg.de).  

\begin{acknowledge}
This work is based on a Core Programme pointed observation (PI:
S. Yu. Sazonov) with \integral, an ESA project with instruments
and science data centre funded by ESA member states (especially the PI
countries: Denmark, France, Germany, Italy, Switzerland, Spain), Czech
Republic and Poland, and with the participation of Russia and the
USA. The authors thank Mikhail Revnivtsev and Eugene Churazov for help
in the data analysis, and Shri Kulkarni for useful suggestions.
\end{acknowledge}

\newpage

\begin{figure}
\caption{\small
The temporal profile of GRB 031203 and its evolution.  {\bf Top --
a}. The profile in the 20--200\,keV energy range obtained with the
\ibis/\isgri\ detector on board \integral.  The binning
is  0.5\,s. Time is measured relative to burst trigger.  A background
level (136\,count\,s$^{-1}$) was estimated from a 200\,s interval preceding 
the trigger and then subtracted from the profile. Vertical error bars indicate
Poissonian noise.  The profile can be classified as a FRED (``Fast
Rise Exponential Decay'') with a rise time of about 1\,s and an
$e$-folding time of $8\pm 0.5\,$s. The peak flux is
$2.6$\,photon\,cm$^{-2}$\,s$^{-1}$ corresponding to $2.4\times
10^{-7}$\,erg\,cm$^{-2}$\,s$^{-1}$ in the 20--200\,keV band.
Two X-ray sources present in the field
of view (Vela\,X-1 and 4U\,0836$-$429) contribute only $\sim 15\,{\rm
count\,s}^{-1}$ to the total count rate (before background subtraction).
Imaging analysis of \ibis\ data revealed no source at the
position of \grb\ during half an hour before nor one day after the
burst. The corresponding 3$\sigma$ upper limits (in the 20--200\,keV
band) are $\sim 10^{-9}\,{\rm erg\,cm}^{-2}{\,\rm s}^{-1}$ and $\sim
10^{-10}\,{\rm erg\,cm}^{-2}\,{\rm s}^{-1}$, respectively.
{\bf Bottom -- b.}
The evolution of the photon index across the duration of the burst.
The spectrum over 20--200\,keV is fitted to a single power
law with index $\alpha$; the bin width varies from 2.5\,s
near the burst peak to 20\,s during the decay phase.
}
\label{fig:LightCurve}
\end{figure}

\begin{figure}
\caption{\small
Spectral energy distribution of \grb\, shown in $\nu
F_\nu$ units. The data points in the 17--500\,keV range were obtained from the
data of the \ibis/\isgri\ detector for the first 20\,s of the
burst, when 80\% of its total energy was emitted. Scattering and
absorption in the interstellar medium of our Galaxy and the host
galaxy of GRB 031203 has negligible effect ($<1$\%) on observed flux
at photon energies above 20\,keV. Vertical bars indicate 1-$\sigma$
statistical uncertainties. We considered a single power law model (photon
index, $\alpha$). Our method \cite{lmr04,rsv+04} consisted of
constructing images in predefined energy intervals followed by
normalizing the resulting source fluxes to the corresponding fluxes of
the Crab nebula for a similar position in the field of view. Analysis
of an extensive set of Crab calibration observations has shown that
the source absolute flux can be recovered with an accuracy 
of 10\% and the systematic uncertainty of relative flux measurement in
different energy channels is less than 5\%. The latter uncertainty was
included in the modelling of the spectrum. The best fit power law model with 
$\alpha=-1.63\pm 0.06$ (1$\sigma$ uncertainty, $\chi^2/{\rm 
dof}=14.8/15$) is shown by the line. We also considered a double power
law model (the so-called ``Band'' model\cite{mgi+82,bmf+93}). Setting
the high energy power law index to $-2$ we are able to place a lower
limit to the peak energy, $E_{\rm peak}>190\,$keV (90\% confidence level).
The cross towards the top left corner of the figure is the
soft X-ray (0.7--5\,keV) fluence, $F_X$ inferred\cite{whl+04,vwo+04} from
the dust scattered halos discovered in {\sl XMM-Newton}
observations. 
}
\label{fig:Spectrum}
\end{figure}

\begin{figure}
\caption{\small
Spectral lag vs. luminosity for cosmological and low-redshift
GRBs. The data for cosmological bursts (unlabeled points), with
measured redshifts ranging between $z=0.84$ and 4.5, are adopted from
ref.~\pcite{s03}. The lag is defined between the burst profiles at 25--50\,keV
and 100--300\,keV, and the luminosity is calculated over the
50--300\,keV band at the burst peak, assuming isotropic emission. Also
plotted are the data for GRB\,980425 ($z=0.0085$)\cite{sdb01}, and \grb\
($z=0.1055$, this work). In the latter case, the peak
luminosity measured with {\sl INTEGRAL} was converted to the 50--300\,keV
range, and the lag (shown with 1$\sigma$ error bars) was determined
between the 20--50\,keV and 100--200\,keV bands. All the lags have been
corrected for cosmological time dilation. The luminosities are given
for a cosmology with $(H_0, \Omega_{\rm m}, \Omega_\Lambda)=
(65$\,km\,s$^{-1}$\,Mpc$^{-1}$, 0.3, 0.7).
}
\label{fig:Lag-Luminosity}
\end{figure}
 
\setcounter{figure}{0}

\begin{figure}
\centerline{\psfig{file=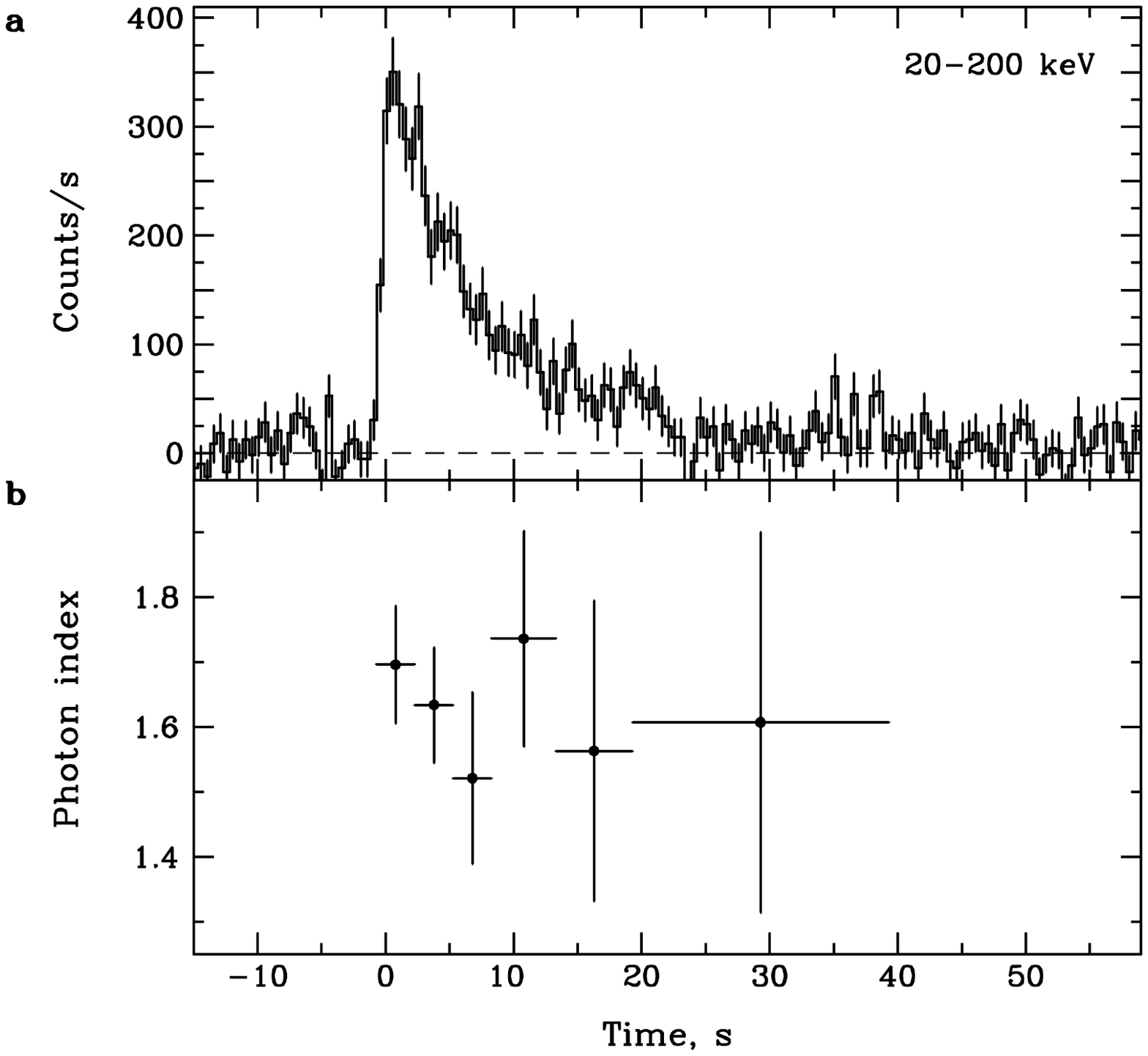,width=\textwidth}}
\caption{}
\end{figure}

\begin{figure}
\centerline{\psfig{file=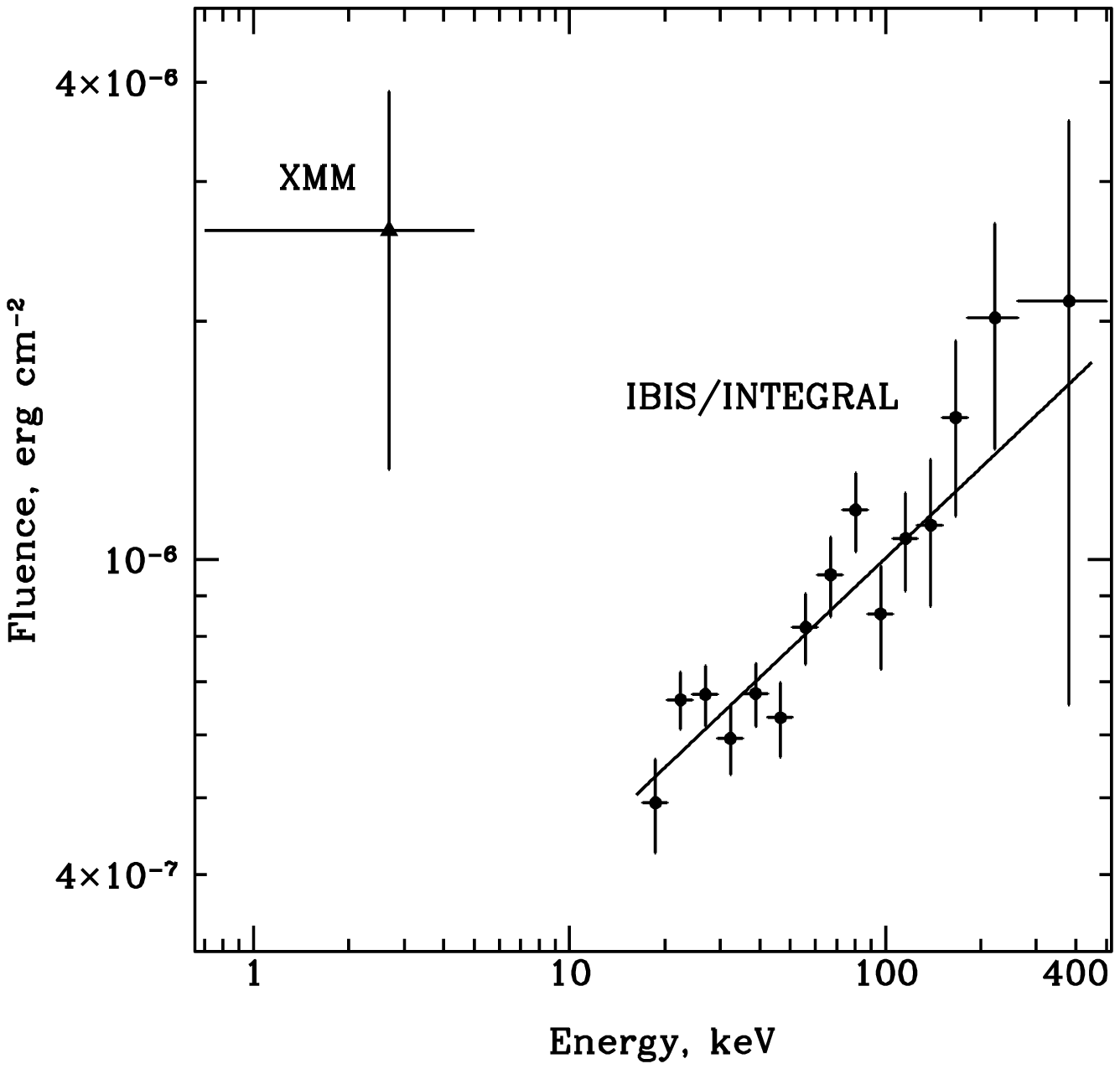,width=\textwidth}}
\caption{}
\end{figure}

\begin{figure}
\centerline{\psfig{file=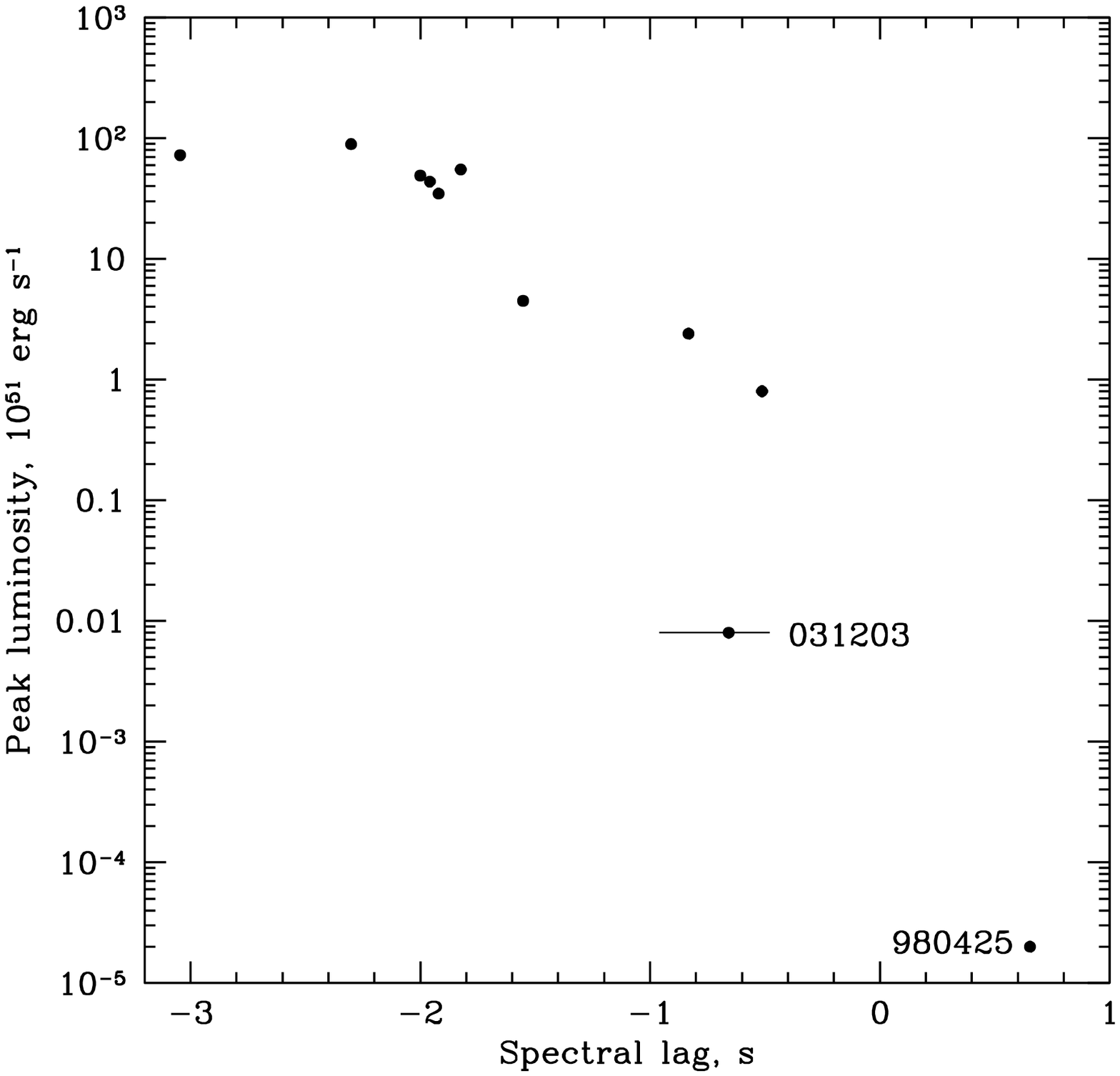,width=\textwidth}}
\caption{}
\end{figure}
 
\end{document}